# Microwave-assisted hydrothermal synthesis of $NH_4V_3O_8$ microcrystals with controllable morphology

G. S. Zakharova[1,2], A. Ottmann[2,*], B. Ehrstein[2], and R. Klingeler[2,3]

[1]Institute of Solid State Chemistry, Ural Division, Russian Academy of Sciences, Pervomaiskaya ul. 91, Yekaterinburg, 620990 Russia

[2]Kirchhoff Institute for Physics, Heidelberg University, INF 227, 69120 Heidelberg, Germany

[3]Centre for Advanced Materials (CAM), Heidelberg University, INF 225, 69120 Heidelberg, Germany

[*] Corresponding author: alex.ottmann@kip.uni-heidelberg.de

**Abstract**

Water-free $NH_4V_3O_8$ microcrystals have been successfully synthesized by a microwave-assisted hydrothermal method. The products were characterized by means of X-ray diffraction, scanning electron microscopy, Fourier transform infrared spectroscopy, thermal gravimetric analysis, cyclic voltammetry, and galvanostatic cycling. The results show phase-pure products whose particle size and morphology can be tailored by varying the reaction conditions, i.e., reaction temperature, synthesis duration, and initial pH value. For instance, at low pH (2.5 – 3), flower-like agglomerates with primary particles of 20 – 30 μm length are found, while at pH = 5.5 single microplates with hexagonal outline (30 – 40 μm) prevail. The sample with the comparably highest specific surface area (11 $m^2/g$) was studied regarding its electrochemical performance. It shows an extraordinary initial discharge capacity of 378 mA h $g^{-1}$ at 10 mA $g^{-1}$, which corresponds to the intercalation of 4.2 $Li^+$/f.u.

# 1. Introduction

Secondary lithium batteries using a lithium-containing metal oxide as the active cathode material and a carbonaceous anode material have been widely used in portable electronics. $LiCoO_2$ is still a principal cathode material for commercial application in Li-ion batteries [1–3]. However, it suffers from many disadvantages such as toxicity, high cost, and irreversibility at elevated temperatures. Compounds based on vanadium oxides have been identified as promising alternative host structures for lithium intercalation [4–6]. Due to the wide range of oxidation states of vanadium from +2 up to +5, these compounds carry the potential to accommodate several Li-ions per formula unit [7,8]. Among them, lithium trivanadate $LiV_3O_8$ is one of the most studied materials for application in Li-ion batteries as it shows high specific capacity and good structural stability [8–10]. Liu et al. [11] fabricated $LiV_3O_8$ which demonstrated the best insertion capacity of 347 mA h $g^{-1}$ (3.7 $Li^+$/f.u.) with excellent cyclic stability. However, the electrochemical performance of $LiV_3O_8$ strongly depends on the synthesis conditions and on post-synthesis thermal treatment. Even worse, it exhibits poor rate performance [12,13]. Replacing $Li^+$ by other cations with larger ionic radii, such as $NH_4^+$ or $Na^+$, but preserving the monoclinic structure, could improve the electrochemical performance of trivanadates [14].

Various methods have been developed to synthesize ammonium trivanadate $NH_4V_3O_8$. The traditional method is to use raw $NH_4VO_3$ as vanadium source, and a mineral acid (HCl or $H_2SO_4$) to adjust the initial pH value and to accomplish an acid hydrolysis [15,16]. Hydrothermal synthesis is a more efficient method to produce $NH_4V_3O_8$ [17]. One benefit is the variety of additional agents (sodium dodecyl sulfonate, sodium dodecyl benzene sulfonate, urea) which can be used to obtain different morphologies and, e.g., nano-scaled materials [18–20]. Recently, microwave-assisted routes have been widely used to synthesize nanoparticles of various oxide compounds [21–23], also in the case of $NH_4V_3O_8$ [24,25]. Compared to the conventional hydrothermal method, microwave synthesis is faster and more energy efficient, while allowing to control the size and shape of the obtained nano- and micro-structures [26,27].

In the present study, we report a simple microwave-assisted method to produce ammonium trivanadate powders. The influence of the preparation conditions on the morphology, size and electrochemical properties of $NH_4V_3O_8$ are examined in detail.

## 2. Experimental

Ammonium metavanadate ($NH_4VO_3$) and concentrated acetic acid were utilized for the synthesis, which proceeds as follows: 40 mg of $NH_4VO_3$ (Sigma-Aldrich) was first dissolved in 40 ml of deionized water to form a clear yellow solution. The desired amount of acetic acid (VWR Chemicals) (corresponding to a final pH value adjusted between 2.5 and 5.0) was added dropwise under stirring to the solution. The obtained solution was poured into a sealed glass vial and then transferred into a microwave reactor (*Monowave 300*, Anton Parr). Within a ramping time of 10 min the vial was heated to a fixed temperature (140, 160, 180, 200, or 220 °C) and held at this temperature for different periods of time. The dwell time was changed between 1 min and 20 min. The products were separated by centrifugation and washed with deionized water several times. For the purpose of comparison, $NH_4V_3O_8$ microcrystals were also prepared by conventional hydrothermal heating in an autoclave according to [17].

X-Ray diffraction (XRD) patterns were obtained from a Shimadzu XRD-7000S using Cu $K_\alpha$ (λ = 1.540 Å) radiation with a step size of $\Delta 2\theta = 0.02°$. The SEM pictures were taken on a ZEISS *Leo 1530* scanning electron microscope. FT-IR spectra were recorded using a *Spectrum One B* (Perkin Elmer). Differential scanning calorimetry (DSC-TG) was performed on a *DSC Q10* (TA Instruments) while heating the samples with 10 K·min$^{-1}$ from room temperature up to 800 °C in air. The specific surface area and pore volume of the samples was measured by a surface area and porosity analyzer (*Gemine VII*, Micromeritics).

For electrochemical characterization (cf. [14]), pristine $NH_4V_3O_8$ powder was stirred overnight with 5 wt% polyvinyldeneflourid binder (Solvay Plastics) and 15 wt% carbon black (Timcal) in N-methyl-2-pyrrolidone (Sigma-Aldrich). The resulting slurry was spread on Al-meshes, dried at ~80 °C under vacuum, mechanically pressed and dried again. The Swagelok-type two-electrode cells were assembled in an Ar-atmosphere glove box. The working electrode and a lithium metal foil counter electrode were separated by two layers of glass fiber separator (Whatman *GF/D*). The electrolyte was 1 mol/l $LiPF_6$ in a mixture of ethylene carbonate and dimethyl carbonate (Merck Electrolyte *LP30*). Electrochemical measurements by means of cyclic voltammetry and galvanostatic cycling with potential limitation (GCPL) were carried out by a *VMP3* Potentiostat System (BioLogic) in the potential range of 1.0 V – 4.0 V vs. Li$^{0/+}$ at constant temperature (25 °C).

## 3. Results and discussion

The structure of the as-prepared microcrystals was studied by means of X-ray powder diffraction. In Fig. 1, exemplary XRD patterns of two samples synthesized under the same conditions (pH = 4, 140 °C for 15 min (MW) / 48 h (AC)) by the microwave-assisted (MW) and the conventional (AC) route, respectively, are shown. All diffraction peaks can be indexed in the monoclinic lattice system with space group *P 21/m* and correspond to $NH_4V_3O_8$ (JCPDS card no. 088-1473 [28]). Lattice parameters were determined by means of full-profile analyses with the FullProf program suite. The MW sample exhibits a = 4.989(4) Å, b = 8.407(5) Å, c = 7.853(4) Å, and β = 96.41(5)° which is in agreement with the lattice constants of the AC one [14]. High purity of both samples is indicated by the absence of any characteristic impurity peaks. The patterns of the MW and the AC $NH_4V_3O_8$ however feature one obvious difference, which is the relative intensity of the (00*l*) Bragg peaks. The AC sample shows in comparison with the other reflections much more pronounced (001), (002) and (003) peaks, e.g. the intensity ratio $I_{(001)}/I_{(011)}$ is more than 25 times greater in the AC sample than in the MW sample. This observation can be attributed to preferred orientation of the single microcrystals of the AC sample (Fig. 2d) as opposed to the more randomly oriented agglomerates (Fig. 2b) which result from the microwave-assisted synthesis.

The morphology of the as-prepared $NH_4V_3O_8$ materials was examined by SEM (Fig. 2). The impact of the synthesis conditions (pH value, reaction temperature, heating duration) on the product's morphology was studied by changing one reaction parameter while the other conditions were kept constant. It turned out that the pH value of the precursor solution is the most crucial factor, determining the morphology of the hydrothermally synthesized $NH_4V_3O_8$. In case of the microwave-assisted hydrothermal treatment, the pH value was varied from 2.5 up to 5.5, while the temperature was always set to 140 °C for 20 min. When the pH value equals 2.5 – 3.0, the dominant microstructures are flower-like three-dimensional (3D) agglomerates composed of many leaf-like microsheets with lengths of 20 – 30 μm (Fig. 2a). The inset (Fig. 2a) reveals that the thickness of the leaf-like particles is in the range of 0.5 – 1.5 μm and that the width is around 10 μm. These microleaves alternatingly overlie each other similar to petals forming a flower. Upon increasing the pH value from 3.0 to 5.5, the morphology gradually transforms from more loosely aggregated microleaves to single microplates with hexagonal outline (Fig. 2b-c). The morphology of the sample obtained at pH = 4 (140 °C) exhibits hexagonal plate-like shape with 0.5 – 1.5 μm thickness, 20 – 30 μm width and 30 – 60 μm length. Some of these microplates can still be found in agglomerates similar to a flower with "open petals" (Fig. 2b). At pH = 5.5, regular hexagonal microplates are observed (Fig. 2c). The dimensions of these hexagons are in the range of 30 – 40 μm, and with 140 - 700 nm they are thinner than the primary particles found at lower pH. The observed differences in morphology can be ascribed to different nucleation rates of $NH_4V_3O_8$ induced by both the different pH values (i.e. $H^+$ concentration) and the different oxidation states of the vanadium ions in the reaction solution [16]. In comparison, the average lengths and thicknesses of the $NH_4V_3O_8$ microplates synthesized by conventional hydrothermal treatment at pH = 4.5 - 5.5 (Fig. 2d) are larger than those of the MW materials. The former show lengths of 60 – 80 μm, widths of 30 – 60 μm, and thicknesses ranging from 1.1 to 2.3 μm.

Varying the temperature of the microwave heating from 140 °C up to 220 °C at fixed pH = 4 and dwell time (20 min) hardly changes the morphology but reduces the particle size ranging from 30 – 60 μm (140 °C) to 20 – 40 μm (220 °C). Concomitantly, the degree of agglomeration decreases as well. Furthermore, it was found that the variation of the dwell time affects the size of the $NH_4V_3O_8$ microcrystals. The corresponding experiments were carried out at pH = 2.5 and 140 °C. The shortest dwell time of 0.5 min yields mainly leaf-like structures with a length of 25 – 30 μm, width of 15 – 20 μm, and thickness of 400 – 900 nm (Fig. 2e). Prolonging the reaction time to 1 min (Fig. 2f) and to 20 min (Fig. 2a), respectively, leads to single leaves crossing each other and further to the flower-like structures described above. At the same time, the thickness and the width of the microleaves increase while the length does not change.

The FT-IR spectrum of $NH_4V_3O_8$, synthesized at pH = 4 and 140 °C for 20 min inside the microwave reactor, is shown in Fig. 3. There are several distinct absorption bands at 3217, 1404, 1006, 966, 737, 597, 528 and 452 cm$^{-1}$, which can be assigned to different excitations of the ammonium trivanadate structure. The bands at 1006 cm$^{-1}$ and 966 cm$^{-1}$ are assigned to V=O symmetric stretching vibrations in the distorted octahedron and the distorted square pyramids, respectively. The band at 737 cm$^{-1}$ corresponds to an asymmetric stretching vibration of the V–O–V bridges. The absorption band centered around 528 cm$^{-1}$ is due to symmetric stretching vibrations of V-O-V bonds. The band at 1404 cm$^{-1}$ is attributed to symmetric bending vibrations of the $NH_4^+$ group. The corresponding asymmetric stretching vibrations are reflected in the peak at 3217 cm$^{-1}$. The pronounced broadening of this stretching band indicates the presence of a dense network of hydrogen bonds [17]. The characteristic band for crystal water around 1620 cm$^{-1}$ [15] is absent, which provides evidence for the pristine material being water-free $NH_4V_3O_8$. This could not be achieved by using the conventional hydrothermal route [17].

In order to confirm the synthesis of water-free $NH_4V_3O_8$ by microwave-assisted hydrothermal synthesis, and to determine the thermal stability of the sample, DSC-TG analysis of the as-prepared material (pH = 3.0, 160 °C for 20 min) was carried out (see Fig. 4). The data display an endothermic peak at 334 °C with an integrated weight loss of about 8.7%, which can be attributed to the decomposition of $NH_4V_3O_8$ according to the following equation:

$$2NH_4V_3O_8 \rightarrow 3V_2O_5 + 2NH_3\uparrow + H_2O \qquad (1)$$

The endothermic feature in the DSC curve at 682 °C demonstrates the melting process of vanadium pentoxide.

During the microwave-assisted hydrothermal process, $NH_4VO_3$ reacts with acetic acid to form ammonium trivanadate molecules in the same way as during the conventional heating [17]:

$$3NH_4VO_3 + CH_3COOH \rightarrow NH_4V_3O_8 + NH_3 + CH_3COONH_4 + H_2O \qquad (2)$$

Comparing the two synthesis methods, the microwave-assisted one permits to obtain crystalline $NH_4V_3O_8$ powder in shorter time and with smaller particle dimensions. The measured specific surface areas are in accordance with the SEM investigations. The specific surface areas of the MW and of the AC sample (pH = 2.5, 140 °C for 10 min / pH = 4, 140 °C for 48 h) are 2.74(3) m$^2$/g and 1.77(3) m$^2$/g, respectively.

The electrochemical performance of a MW sample (pH = 2.5, 220 °C for 0.5 min) with a comparatively high specific surface area of 11(1) m$^2$/g was studied by means of cyclic voltammetry and galvanostatic cycling, both in the range of 1.0 – 4.0 V vs. Li$^{0/+}$. The results are depicted in Fig. 5 and are compared to those of conventionally synthesized microbelts (Fig. 2d), which were discussed in detail elsewhere [14]. The first cycle of the cyclic voltammograms (CV) (Fig. 5a) shows that both samples exhibit the same redox features with two dominant peaks around 1.6 V (reduction) and 3 V (oxidation), respectively. The redox activity originates from the V$^{5+/4+}$ and V$^{4+/3+}$ couples, and is associated with the intercalation and deintercalation of Li$^+$-ions into the layered NH$_4$V$_3$O$_8$ structure [14,18]. The MW sample features noticeably higher current intensities in the CV, which can be ascribed to the greater active surface area (11 vs. 1.77 m$^2$/g). This effect carries over to the initial charge/discharge capacities during the GCPL measurement at 10 mA g$^{-1}$. The MW sample achieves an initial discharge capacity of 378 mA h g$^{-1}$, which corresponds to the intercalation of 4.2 Li$^+$/f.u., compared to 261 mA h g$^{-1}$ (2.9 Li$^+$/f.u.) in case of the AC microbelts. However, during the further course of the GCPL with increasing charge/discharge rates of 20, 50, 100 and 200 mA h g$^{-1}$, respectively, the capacities converge and no significant difference can be observed anymore. In contrast to the microbelts [14], the MW sample shows decreasing capacities during the first seven cycles which could be caused by irreversible structure changes/damage due to too deep discharge (Li$^+$ intercalation).

## 4. Conclusions

Morphology-controlled synthesis of NH$_4$V$_3$O$_8$ microcrystals was achieved via a microwave-assisted hydrothermal approach with short reaction times (< 30 min). The reaction conditions, in particular the pH value of the precursor solution, show a considerable impact on the particle size and morphology of the synthesized powder materials. The morphologies can be varied from flower-like agglomerates (pH = 2.5, 140 °C for 20 min) to single hexagonal-shaped platelets (pH= 5.5, 140 °C for 20 min) with particle dimensions of the order of 10 μm. XRD, DSC-TG and FT-IR measurements confirm the phase purity of the samples, and the absence of characteristic absorption bands in the FT-IR spectrum indicates that the pristine $NH_4V_3O_8$ is water-free. Electrochemical characterization by means of CV and GCPL reveals typical $Li^+$ de-/intercalation activity attributed to the $V^{5+/4+}$ and $V^{4+/3+}$ redox couples with a high initial discharge capacity of 378 mA h $g^{-1}$ at 10 mA $g^{-1}$.

**Acknowledgement**

This work was supported by the Ministry of Science and Education of Russian Federation (unique project identifier RFMEF161314X0002) and by the CleanTech-Initiative of the Baden-Württemberg-Stiftung (Project CT3: Nanostorage). G.S.Z acknowledges support by the Excellence Initiative of the German Federal Government and States. A.O. acknowledges support by the IMPRS-QD.

**Figure captions**

1) XRD patterns of $NH_4V_3O_8$ prepared by the conventional hydrothermal treatment (black line; AC) and the microwave-assisted method (blue line, MW). Inset: Detail of the data with Bragg positions according to JCPDS card no. 088-1473 [28] marked in green.

2) SEM images of $NH_4V_3O_8$ synthesized by (a, b, c, e, f) microwave-assisted method and (d) conventional hydrothermal treatment under different conditions: (a) at 140 °C and pH = 2.5 for 20 min, (b) at 140 °C and pH = 4.0 for 20 min, (c) at 140 °C and pH = 5.5 for 20 min, (d) at 160 °C and pH = 4 for 48 h, (e) at 140 °C and pH = 2.5 for 0.5 min, (f) at 140 °C and pH = 2.5 for 1 min.

3) IR spectrum of $NH_4V_3O_8$ synthesized by microwave-assisted method at 220 °C and pH =2.5 for 0.5 min.

4) DSC and TG scans of $NH_4V_3O_8$ powder synthesized by microwave-assisted method at 160 °C and pH = 3.0 for 20 min (initial sample mass 10.643 mg).

5) (a) Cyclic voltammograms of MW (red line; pH = 2.5, 140 °C for 20 min) and AC (black line; pH = 4, 140 °C for 48 h) materials. (b) Discharge capacities for the same samples at various current rates.

**Figures**

1)

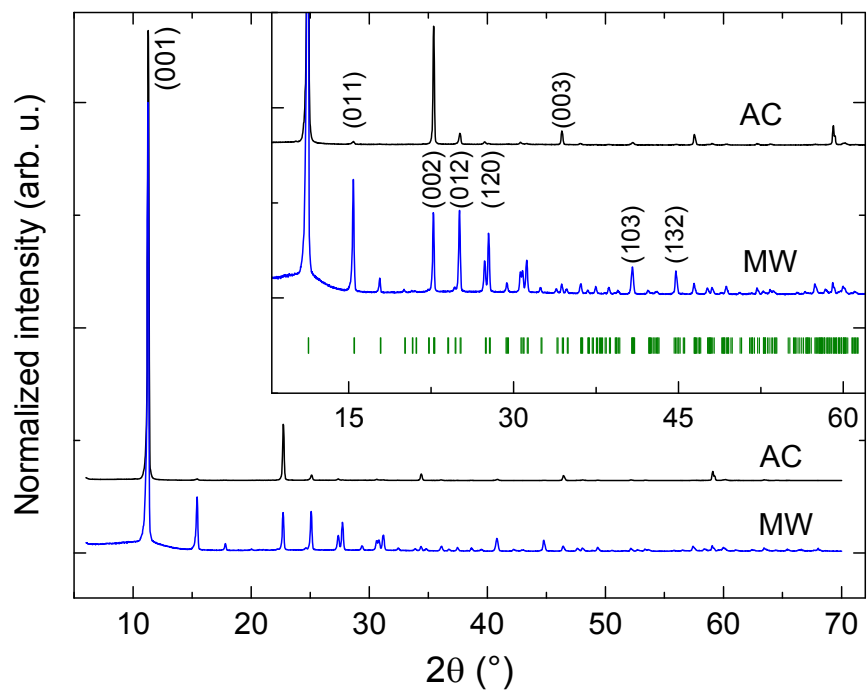

2)

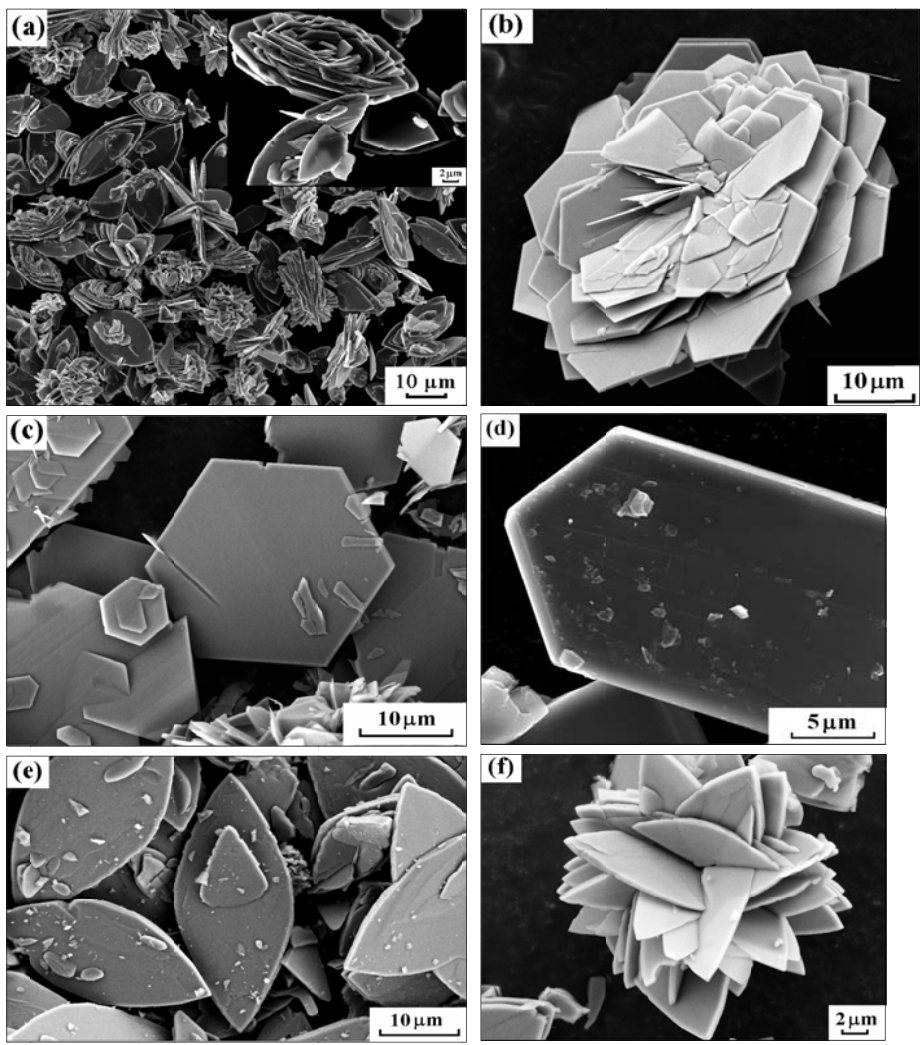

3)

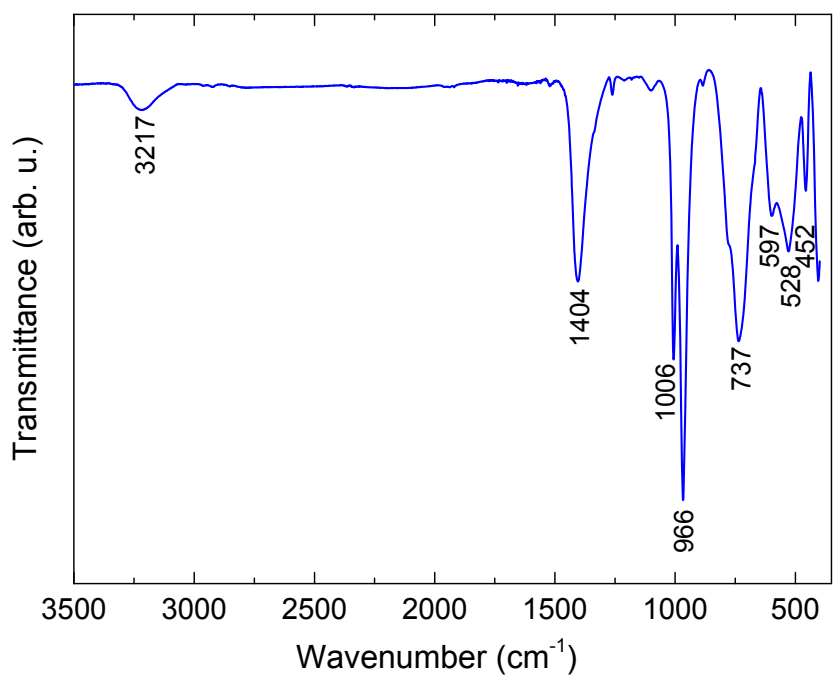

4)

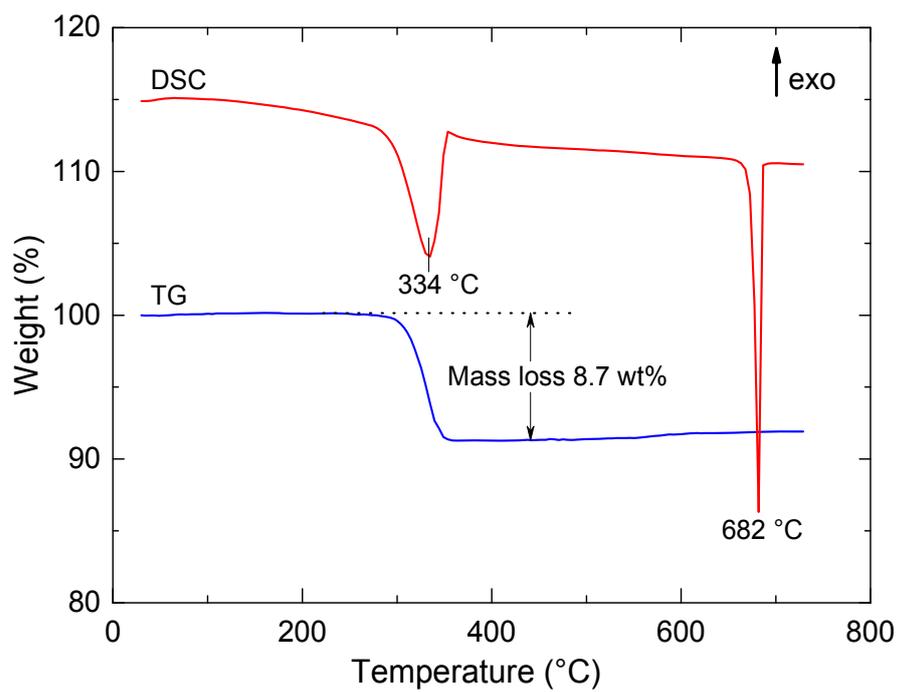

5)

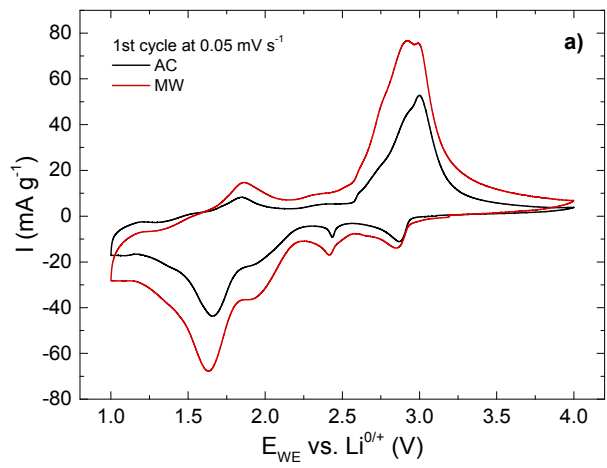 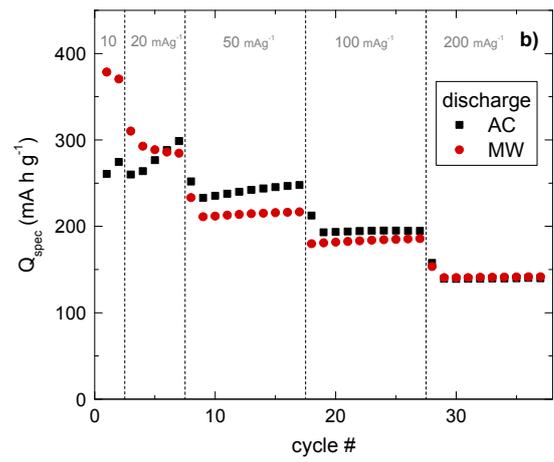